\documentclass[aps,pra,twocolumn,showpacs,superscriptaddress]{revtex4}
\usepackage{graphicx}
\usepackage{amsmath}
\usepackage{amssymb}
\usepackage{color}
\newcommand{\rrho}{\ensuremath{\boldsymbol{\rho}}}
\newcommand{\E}{\mathcal{E}}
\newcommand{\I}{\mathcal{I}}

\graphicspath{{converted_graphics/}}

\begin{document}
\title{Phase conjugation and mode conversion in stimulated parametric down-conversion with orbital angular momentum: a geometrical interpretation}

\author{A. G. de Oliveira}
\affiliation{Departamento de F\'isica, Universidade Federal de Santa Catarina, Florian\'opolis, SC, 88040-900, Brazil}

\author{M. F. Z. Arruda}
\affiliation{Departamento de F\'isica, Universidade Federal de Santa Catarina, Florian\'opolis, SC, 88040-900, Brazil}
\affiliation{Instituto Federal do Mato Grosso, Sorriso, MT, 78890-000, Brazil}

\author{W. C. Soares}
\affiliation{Departamento de F\'isica, Universidade Federal de Santa Catarina, Florian\'opolis, SC, 88040-900, Brazil}
\affiliation{N\'ucleo de Ci\^encias Exatas - NCEX, Universidade Federal de Alagoas, Arapiraca, AL, 57309-005, Brazil}

\author{S. P. Walborn} 
\affiliation{Instituto de F\'isica, Universidade Federal do Rio de Janeiro, Caixa Postal 68528, Rio de Janeiro, RJ, 21945-970, Brazil}

\author{A. Z. Khoury} 
\affiliation{Instituto de F\'isica, Universidade Federal Fluminense, Niter\'oi, RJ, 24210-346, Brazil}

\author{A. Kanaan}
\affiliation{Departamento de F\'isica, Universidade Federal de Santa Catarina, Florian\'opolis, SC, 88040-900, Brazil}

\author{P. H. Souto Ribeiro} 
\affiliation{Departamento de F\'isica, Universidade Federal de Santa Catarina, Florian\'opolis, SC, 88040-900, Brazil}

\author{R. Medeiros de Ara\'ujo}
\affiliation{Departamento de F\'isica, Universidade Federal de Santa Catarina, Florian\'opolis, SC, 88040-900, Brazil}

\date{\today}
\begin{abstract}
We report on an experiment that investigates the spatial mode conversion in the process of parametric down-conversion seeded by a light beam in a superposition of orbital angular momentum modes. This process is interpreted in terms of a geometric representation of first-order spatial modes in a Poincar\'e sphere, providing an intuitive image of the phase conjugation and the  topological charge conservation. We also make a comparison with the analogous phenomenon for optical parametric oscillators.
\end{abstract}
\maketitle
\section{Introduction}

Parametric down-conversion (PDC) is a nonlinear optical process that has revolutionized the field of Quantum Optics \cite{zeilinger99,walborn10a,pan12} and an important experimental tool for investigating quantum entanglement and its applications in Quantum Information and Computation \cite{walther05}. This process can be understood as the coupling between three optical modes --- pump ($p$), signal ($s$) and idler ($i$) --- via a $\chi^{(2)}$ interaction inside a nonlinear crystal \cite{saleh91}. One pump photon is converted into two photons, usually called signal and idler, in an almost elastic process, so that energy ($\omega_p = \omega_s + \omega_i$) and momentum ($\vec{k}_p = \vec{k}_s + \vec{k}_i$) are conserved (see Figure \ref{fig:stimPDC}a). This system presents several quantum signatures, in the sense that there is no classical counterpart for its behavior. The emission of signal and idler photons can occur spontaneously, which cannot be explained by classical nonlinear optics. Moreover, the simple fact that one measures coincidences between  signal and idler photons arriving to two different photon counters with a high enough coincidence counting rate indicates that a classical inequality is violated \cite{mandel91}. Other quantum signatures stem from the correlations between transverse components of the wavevectors of the signal and idler photons \cite{walborn10a}. In addition to the quantum properties that can be observed in sources of parametric down-conversion without any special arrangement, one can also prepare and measure entanglement in other degrees of freedom, such as polarization. Polarization-entangled photon pairs can be prepared using a few different schemes \cite{kwiat95,kwiat99} and are one of the most important experimental resources in Quantum Information Science.

The parametric down-conversion process can be stimulated in at least two ways. One method consists in placing the crystal inside an optical cavity, resulting in the so-called optical parametric oscillator (OPO) \cite{heidmann87}. This device emits light beams that are quantum correlated in their intensities and phases \cite{villar05,barbosa18} and has also found crucial applications in Quantum Information Science \cite{kimble08}. Another way of stimulating parametric down-conversion is by simply aligning an auxiliary laser with one of the down-converted beams \cite{wang91}, as illustrated in Figure \ref{fig:stimPDC}b. When spontaneous emission takes place, there is a very broad range of accessible modes in which the emission occurs. In the presence of the auxiliary laser field, a narrower range of modes is privileged by stimulated emission. This stimulation in the signal direction also affects the emission rate in its counterpart, the idler, which is indirectly intensified. We refer to this process as stimulated parametric down-conversion (StimPDC). 

The intensified light emitted by StimPDC has been studied from the perspective of its coherence properties \cite{wang91} and its transverse spatial characteristics \cite{SoutoRibeiro99}. Most of the physical properties of the stimulated light can be described classically, even though quantum properties appear if proper detection schemes are implemented. One example is the photon addition experiment \cite{parigi07}. Also of great interest are the transverse spatial properties of the light beams involved in StimPDC. It is known that images and angular spectra are transferred from the pump to signal and idler in the process \cite{SoutoRibeiro99}, as well as the orbital angular momentum \cite{Caetano02}. Moreover, the phase anti-correlation imposes phase conjugation of the idler beam with respect to the auxiliary stimulating laser, and this has some counterintuitive consequences \cite{caetano01}. 

Here, we perform a StimPDC experiment in which the pump is a Gaussian mode and the signal is seeded by an auxiliary beam prepared in a general superposition of the first-order Laguerre-Gaussian (LG) modes. The set of such superpositions allows for a simple geometrical representation in a Poincar\'e sphere \cite{padgett99}, where each point in its surface corresponds to a specific coherent superpostition of first-order modes. We show that, due to phase conjugation, the points representing the signal and idler modes in the Poincar\'e sphere are the specular reflection of each other with respect to the equatorial plane of the sphere. These results are equivalent to those obtained in references \cite{santos07,rodrigues2018} for an OPO. The mode coupling for the processes with and without a cavity is very different due to the constraints imposed by the cavity mirrors and geometry. We make a comparison between the two processes.

\section{Stimulated down-conversion}
\label{sec:stimPDC}
\begin{figure}
\centering
  \includegraphics[width=\columnwidth]{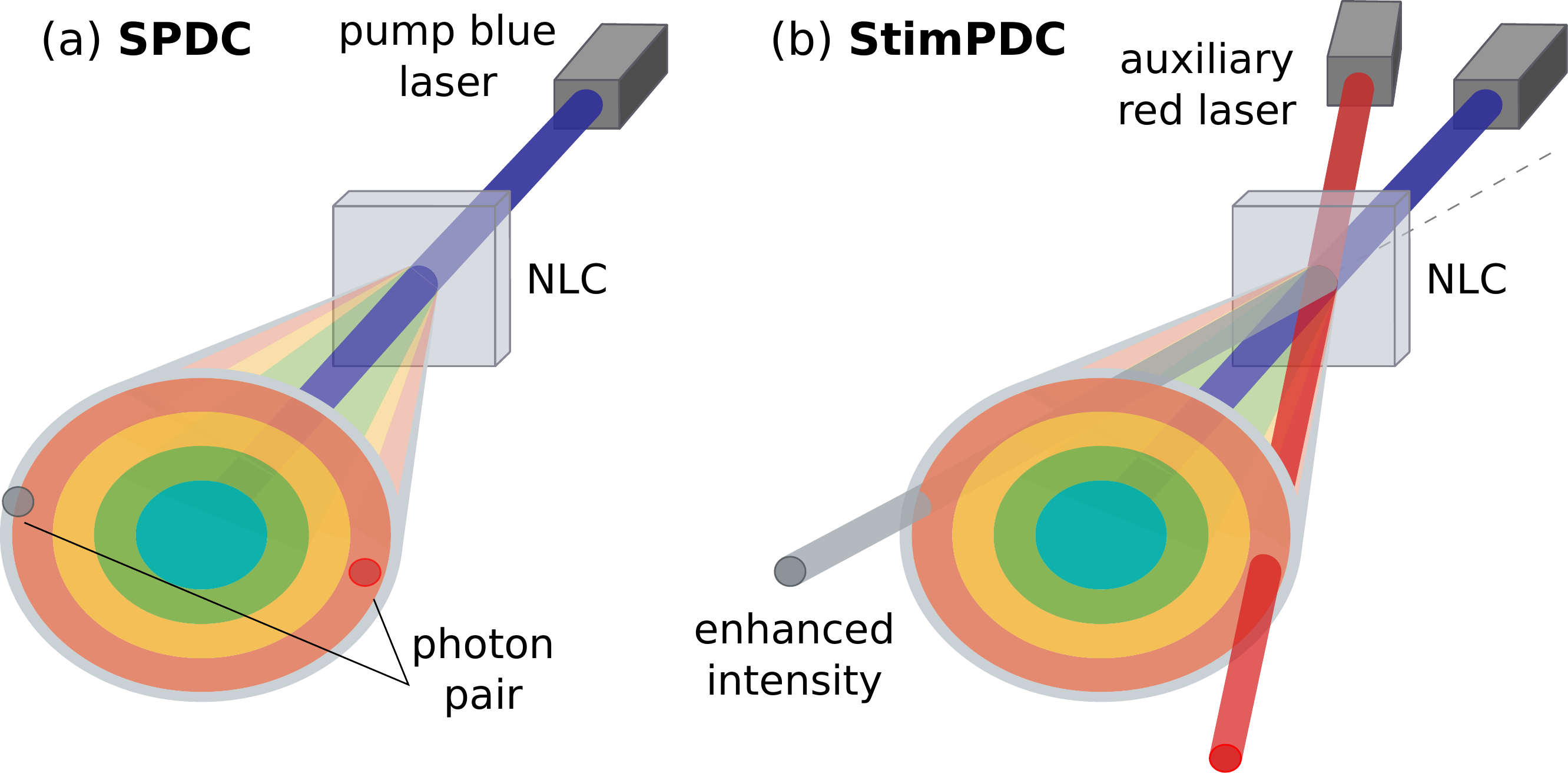}
   \caption{Schemes for (a) spontaneous parametric down-conversion (SPDC) and (b) stimulated parametric down-conversion (StimPDC). In SPDC, a nonlinear $\chi^{(2)}$ crystal (NLC) is pumped by an intense laser, producing pairs of entangled photons in various directions with different wavelengths that form a cone barely visible to the naked eye. In StimPDC, a laser prepared in the appropriate mode (correct wavelength, polarization and direction) interacts with the pump inside the crystal, enhancing the conversion of pump photons into signal and idler photons in a given pair of directions.}
   \label{fig:stimPDC}
\end{figure}

Fig. \ref{fig:stimPDC}b shows the scheme for StimPDC, which needs an auxiliary laser, here also called signal, in addition to the pump and the nonlinear crystal. In reference \cite{SoutoRibeiro99}, the intensity of the idler beam as a function of the transverse position, under the paraxial and thin-crystal approximations, is derived as:
\begin{align}
\I(\rrho_i)&\propto \int d\rrho\ |\E_p(\rrho)|^2 + \nonumber \\
&+ \left| \int d\rrho\ \E_p(\rrho) \E_s^*(\rrho)\ \exp \left(i\,|\rrho_i - \rrho|^2 \dfrac{k_i}{2z}\right)\right|^2 ,
\label{eq:stimPDC}
\end{align}  
where  $\E_p$ ($\E_s$) is the amplitude envelope of the pump (signal) field at the transverse position $\rrho$ in the plane of the crystal; $\rrho_i$ is a vector defining a position in a plane transverse to the propagation of the idler beam, situated at a distance $z$ from the crystal; and $k_i$ is the idler wavenumber.

The first term on the right of Eq. (\ref{eq:stimPDC}) describes the spontaneous emission that is always present in the process, whether it is stimulated or not. The second term describes the stimulated emission component. We can see that it is given by the product between the pump and the complex conjugate of the signal transverse field distributions in the crystal. The result of this product is the idler transverse field distribution, which is freely propagated from the crystal to the detection plane.

The contribution from stimulated emission is often much more significant than the one from spontaneous emission. It suffices that the auxiliary laser is intense, which is typically the case, since a few milliwatts of stimulating laser is enough to make the stimulated emission $\sim 10^2$ times more intense than the spontaneous emission, which may then be neglected.  

Let us suppose that the auxiliary laser  is prepared in an arbitrary superposition of first-order Laguerre-Gauss modes:
\begin{equation}
\E_{s}(\rrho) = \cos\frac{\theta}{2} LG_{+}(\rrho) + e^{i\phi} \sin\frac{\theta}{2}  LG_{-}(\rrho),
\label{eq:superposition}
\end{equation}
where the angle $\theta$ determines the weight of each mode and $\phi$ is the relative phase between them. Here $LG_{\pm}$ refers to the first-order Laguerre Gauss mode with topological charge $\pm1$. Let us also suppose that the pump field is a well collimated and large enough beam so that $\E_p(\rrho)$ can be considered approximately constant. In this case, if we can neglect the spontaneous emission term in Eq. (\ref{eq:stimPDC}), the idler intensity will be:
\begin{align}
\I(\rrho_i)\propto \left|\int d\rrho \right.&
\left[\cos\frac{\theta}{2} LG_+(\rrho)+e^{i\phi}\sin\frac{\theta}{2}  LG_-(\rrho)\right]^* \nonumber \\ 
& \times\left.\exp \left(i\,|\rrho_i - \rrho|^2 \frac{k_i}{2z}\right)\right|^2.
\label{eq:stimonly}
\end{align}  
The above integral corresponds to a free paraxial propagation over a distance $z$ with wavenumber $k_i$. Now, LG modes are invariant under paraxial propagation, except for a scaling factor $a$, which depends on the initial beam waist and on the propagation distance. Therefore, the idler intensity at a plane situated at a distance $z$ from the crystal simplifies to:
\begin{align}
&\I(\rrho_i) \propto \left| \cos\frac{\theta}{2} LG_-(a\rrho_i) + e^{-i\phi} \sin\frac{\theta}{2}  LG_+(a\rrho_i) \right|^2 \nonumber\\ 
&= \left| \cos\frac{\pi-\theta}{2} LG_+(a\rrho_i) + 
e^{i\phi} \sin\frac{\pi-\theta}{2}  LG_-(a\rrho_i) \right|^2\!\!.
\label{eq:conjugation}
\end{align}  
We see that idler remains a superposition of Laguerre-Gauss modes and the modal structure of the field described by Eq. (\ref{eq:conjugation}) resembles that of Eq. (\ref{eq:superposition}): the weights of the positive and negative LG modes are interchanged, but their relative phase remains constant. In summary, under the realistic approximations discussed above, the phase conjugation imposes that:
\begin{align}
\theta_i & =\pi-\theta_s, \label{eq:theta}\\
\phi_i & =\phi_s. \label{eq:phi}
\end{align} 

\section{Geometrical interpretation}
\label{poincare}

\begin{figure}
\centering
  \includegraphics[width=\columnwidth]{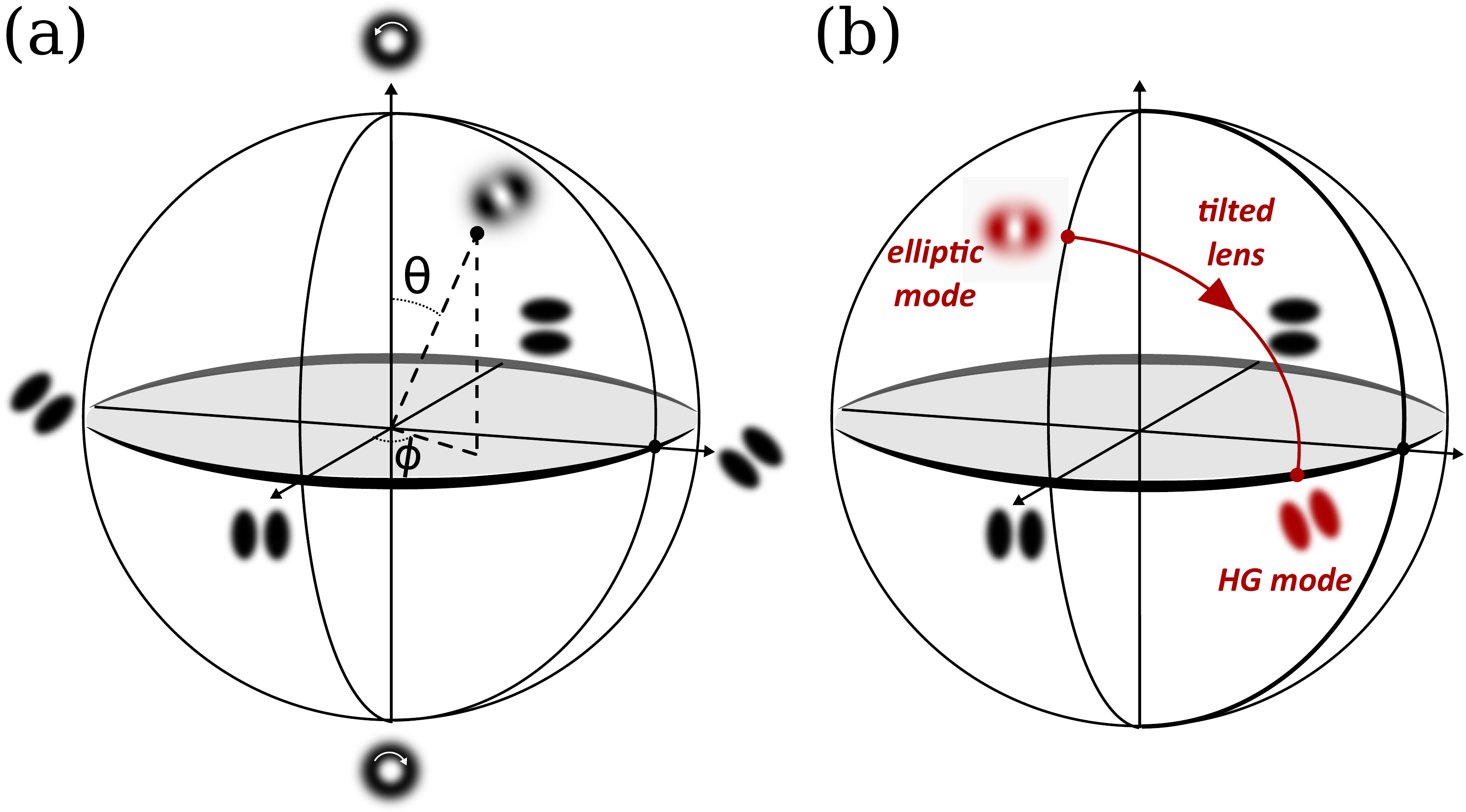}
  \caption{(a) Optical first-order Gaussian modes geometrically represented in a Poincar\'e sphere. (b) Using a horizontally tilted lens, one can apply, to any given mode on the sphere, a $\pi/2$ rotation around the axis defined by Hermite-Gaussian modes TEM$_{10}$ and TEM$_{01}$.}
  \label{fig:sphere}
\end{figure}

Similarly to the Bloch sphere, used as a geometrical representation of the pure state space of a qubit, the Poincar\'e sphere is a useful graphical tool for picturing different types of polarized light. Each point in the sphere represents one type of polarization (among linear, circular and elliptical), so that one can visualize unitary dynamics of the system as a trajectory on its surface. An analogy between polarization modes and Gaussian beams can be made, where linear polarization modes correspond to first-order Hermite-Gaussian modes and circular polarization modes correspond to first-order Laguerre-Gaussian modes \cite{padgett99}. Figure \ref{fig:sphere}a depicts the Poincar\'e sphere for Gaussian modes as described above. The angle $\theta$ in Eq. (\ref{eq:superposition}) now plays the role of the polar angle in spherical coordinates, while $\phi$ corresponds to the azimuthal angle.

This analogy provides a geometrical representation where the trajectories in the sphere are associated with mode conversion. In Ref. \cite{santos07}, the mode conversion was theoretically analyzed for an optical parametric oscillator and interpreted in terms of the Poincar\'e sphere. It is shown that the idler beam is described by a point in the sphere that is the specular reflection of the signal with respect to the equatorial plane. An experimental demonstration of this phenomenon came recently in \cite{rodrigues2018}. Eqs. (\ref{eq:theta}) and (\ref{eq:phi}) show that the same effect can be observed in stimulated down-conversion without a cavity. The main difference between the OPO and StimPDC concerns the mode coupling in the parametric process. While, in StimPDC, only the phase matching conditions and mode overlap rule mode coupling, in the OPO, there are additional constraints imposed by the cavity mirrors and geometry that may suppress or enhance the strength of the coupling.

\section{Experiment}
\label{sec:exp}
\begin{figure}
  \includegraphics[width=\columnwidth]{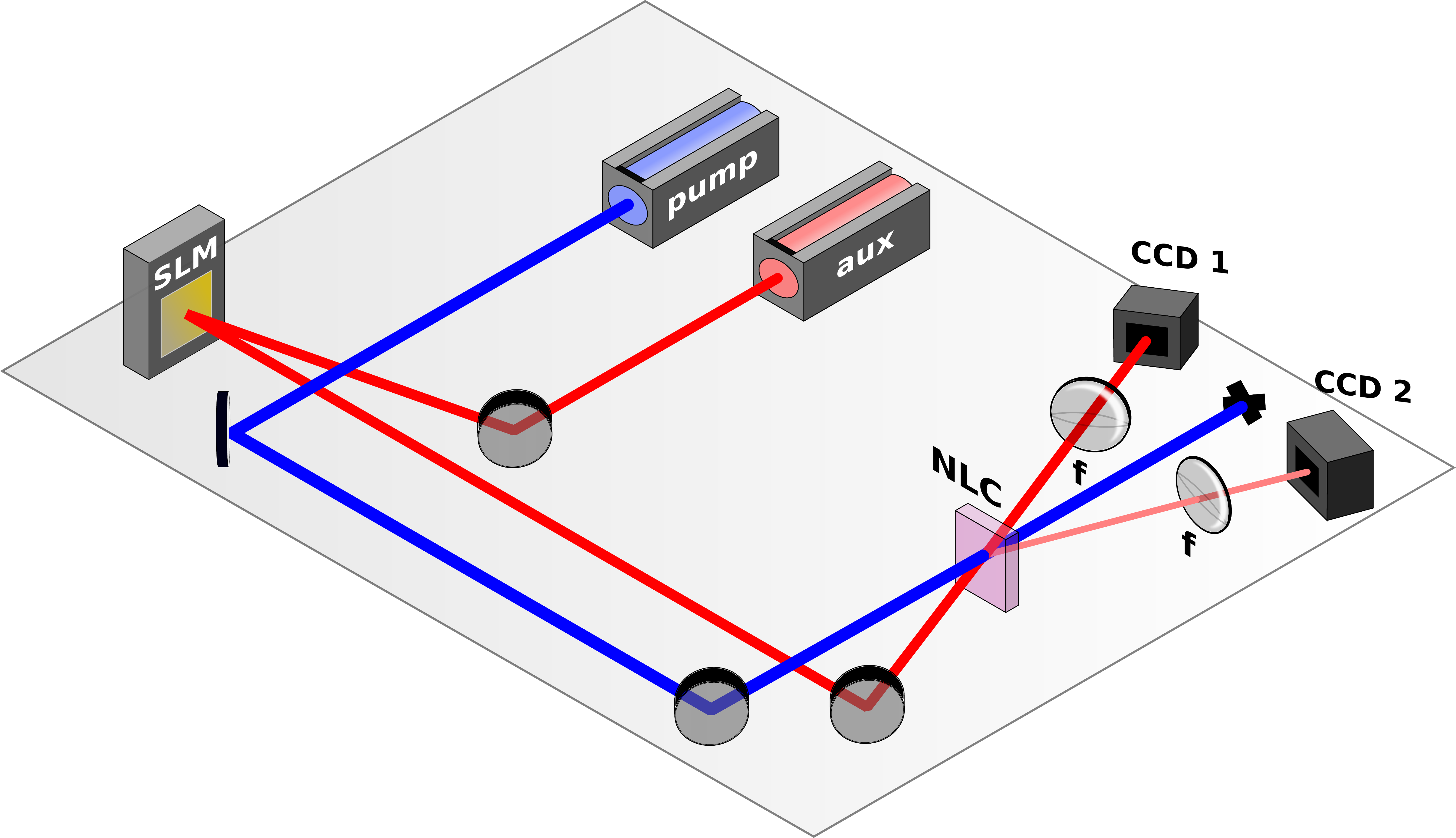}
   \caption{Experimental setup. A 405-nm laser pumps a nonlinear crystal (NLC), while a 780-nm laser (aux) seeds the parametric interaction after being reflected by a spatial light modulator (SLM). Two charge-coupled devices (CCD1 and CCD2) register the intensity profiles of signal and idler beams. Each CCD camera is placed at the focal plane of a spherical lens ($f$).}
   \label{fig:setup}
\end{figure}
Fig. \ref{fig:setup} shows a sketch of the experimental set-up. A 405 nm diode laser is used to pump a BBO crystal and produce parametric down-conversion. Within the spontaneous emission cone, we select signal and idler beams at 780 nm and 840 nm, respectively, using 10-nm bandwidth interference filters prior to detection. The angle between the pump linear polarization and the optical axis of the crystal is chosen in such a way that the directions of propagation of signal and idler beams make an angle of $\simeq 4^\circ$ with the pump beam direction. 

We start counting photons with single-photon counting modules and measuring coincidences between the time of detection of signal and idler photons, in order to identify their respective directions. The correct directions are recognized when the coincidence counting is maximum. The next step is to align the auxiliary laser (a diode laser at 780 nm) along the path of the signal beam. At this point the photon counters are replaced with CCD cameras, so that we can take pictures of the signal and idler transverse profiles. The use of CCD cameras instead of photon counters is possible because the stimulated emission greatly enhances the intensity of the idler beam. This considerably improves the efficiency of the experimental procedure in terms of alignment and measurement time, as compared to experiments where only single-photon counting modules are used.

Before the auxiliary laser is sent to the crystal, it is reflected by a phase-only spatial light modulator (SLM), so that we can prepare its transverse profile with high versatility. We use it to prepare superpositions of modes described by Eq. (\ref{eq:superposition}) to a good approximation, with the ability of varying $\theta$ and $\phi$ on demand. The measurement results are images of the signal and idler profiles.

\begin{figure}
  \includegraphics[width=\columnwidth]{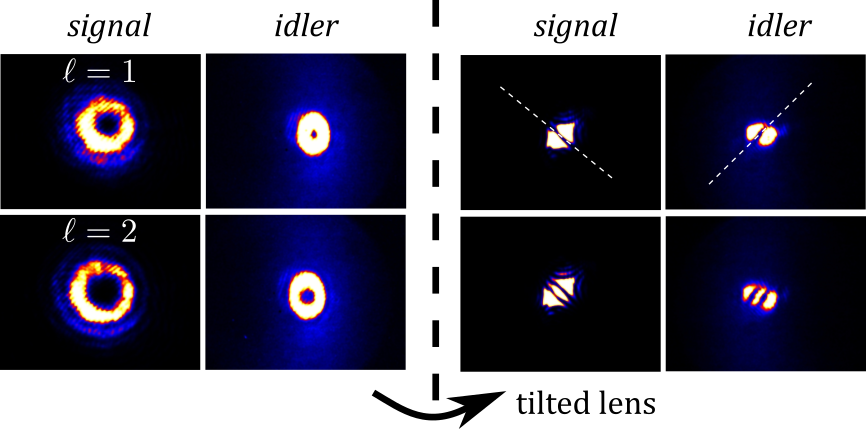}
   \caption{Experimental results showing OAM conservation with the help of a tilted lens, which reveals the sign of the topological charge. The faint vertical blue band on the idler profile is a small arc of the spontaneous emission light cone.}
   \label{fig:oam}
\end{figure}

We first show results illustrating the conservation of orbital angular momentum (OAM) in the StimPDC process and recall the tilted-lens method \cite{vaity2013} for detecting the sign of the topological charge of a mode. In StimPDC, when the auxiliary beam is prepared with topological charge $\ell$, the idler's is $-\ell$ \cite{Caetano02}. Figure \ref{fig:oam} shows the cases $\ell=1$ and $\ell=2$. As we can see, after passing through a tilted lens and undergoing some propagation, a mode with positive (negative) charge acquires the shape of a Hermite-Gaussian mode with dark fringes extended along an axis with a negative (positive) slope, as indicated by the dashed white lines. The number of dark fringes acquired by the beam is equal to its topological charge.

\begin{figure}
  \includegraphics[width=\columnwidth]{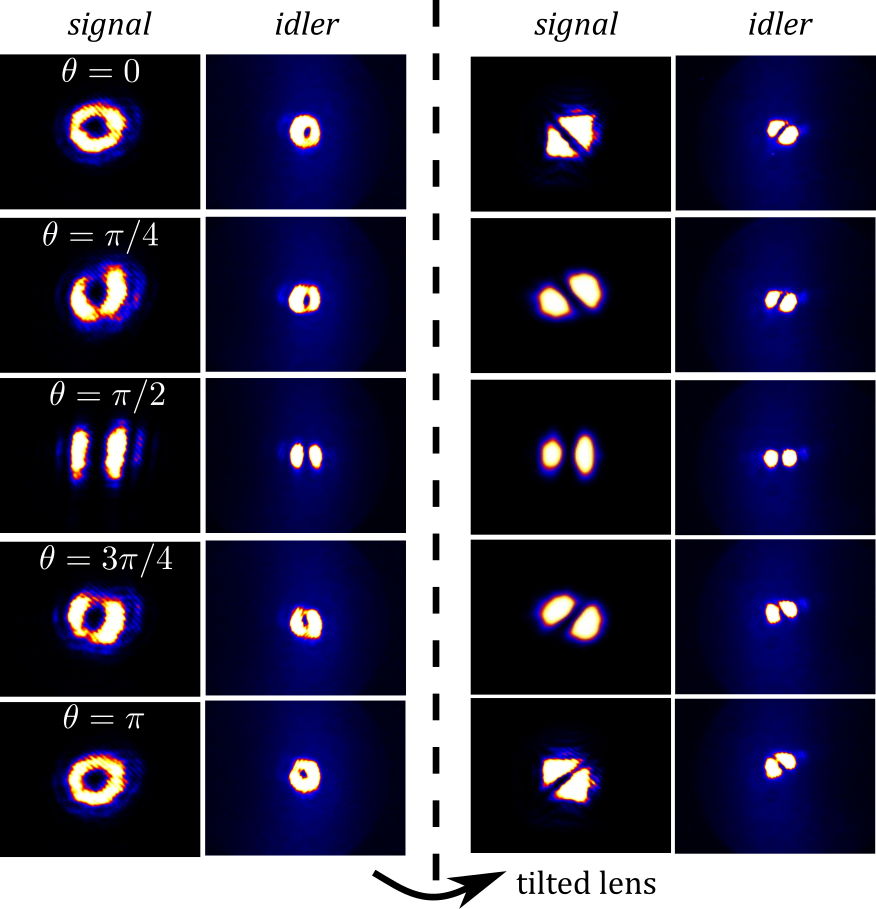}
   \caption{Experimental results along the $\phi=0$ meridian. On the left, intensity profiles of the free propagating beams. On the right, intensity profiles obtained after a tilted lens, showing opposite paths on the Poincar\'e sphere.}
   \label{fig:meridian}
\end{figure}

Within the subspace of first-order Gaussian modes, the action of a horizontally tilted lens has an interesting geometrical interpretation on the Poincar\'e sphere: it rotates any mode by an angle $\pi/2$ around the axis on which lie its two eigenmodes (TEM$_{10}$ and TEM$_{01}$). As a consequence, any mode lying on the $\phi=0$ meridian is mapped by the tilted lens onto the equator of the sphere (see Figure \ref{fig:sphere}b). We have used this property to demonstrate the symmetry described in the previous section for the case of modes along the $\phi=0$ meridian. The experimental results are shown in Figure \ref{fig:meridian}, in which the polar angle of the auxiliary beam is varied from 0 to $\pi$ (North pole to South pole), while we observe the formation of an idler beam going from South pole to the North Pole, at the same pace but in the opposite direction. This shows signal and idler are specular images of each other with respect to the equatorial plane.

\begin{figure}
  \includegraphics[width=\columnwidth]{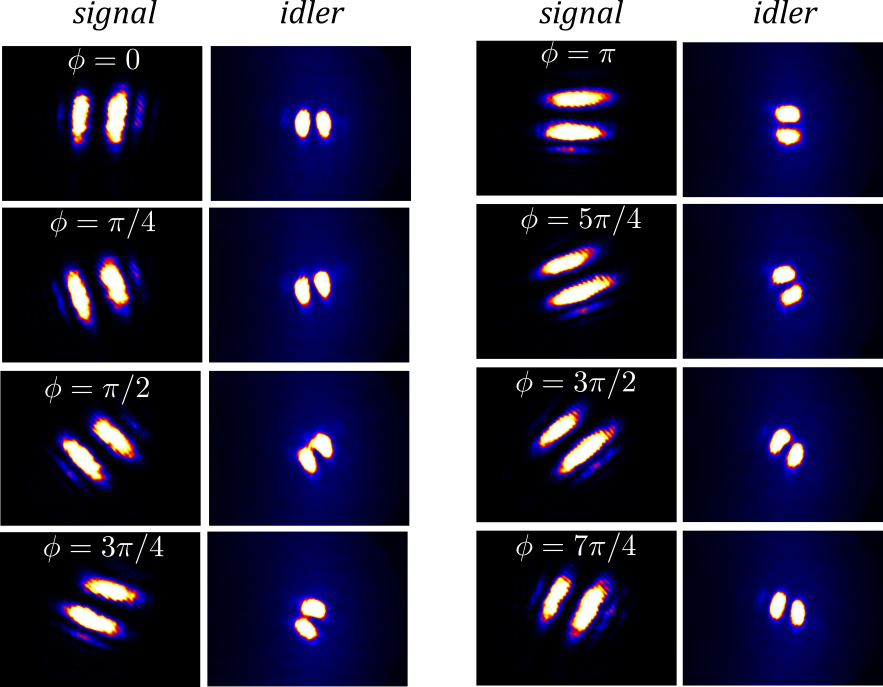}
   \caption{Experimental results around the equator of the Poincar\'e sphere ($\theta=\pi/2$).}
   \label{fig:equator}
\end{figure}

In Figure \ref{fig:equator}, we can see four columns of pictures. The first and third are the profile of the auxiliary laser as captured by the camera when one keeps its polar angle $\theta=\pi/2$ and continuously changes the azimuthal angle $\phi$ from 0 to $2\pi$. The second and fourth columns correspond to the idler beam. We can see that all patterns look like Hermite-Gaussian beams and their symmetry axes rotate as $\phi$ is incremented. This means that the idler beam follows the signal along the Equator, which is another demonstration of the equatorial plane symmetry in StimPDC.

We note that our experimental results confirm what is known concerning conservation of OAM in the generation of LG modes \cite{Mair2001,Walborn04a} and parity conservation of Hermite-Gaussian modes \cite{Walborn2005,Straupe2011} in spontaneous parametric down-conversion.

\section{Conclusion}

In conclusion, we have investigated experimentally the mode conversion in the process of stimulated parametric down-conversion without a cavity. In our analysis, we have prepared superpositions of first-order Laguerre-Gaussian modes with variable weights and relative phase and measured the stimulated beam obtained in the idler mode. We interpret the evolution of the light modes in the Poincar\'e sphere for the first-order Laguerre and Hermite-Gaussian beams in analogy to circular and linear polarization states. The trajectories of the signal and idler beams in the sphere are related by a specular reflection, highlighting the phase conjugation between these two spatial modes imposed by the phase matching in the parametric interaction.

These results can also be interpreted in terms of the Advanced-Wave Picture (AWP) for StimPDC \cite{Arruda2018}, in which the idler wave is the equivalent of the signal advanced wave being reflected by the pump wavefront at the crystal plane. Applying the AWP to our specific case, the idler must be the equivalent of the reflection of the signal by a plane mirror. Indeed, that is exactly what is observed experimentally: idler and signal fields present the same intensity distribution but opposite topological charge. It is important to note that the AWP is also a consequence of phase conjugation.

The tilted lens method was utilized to locate the hemisphere of the modes when considering the trajectory $\phi=0$ on the Poincar\'e sphere. General trajectories may be explored provided that the lens tilt is controlled in other directions, as in \cite{Silva2018}.

Arriving at a similar geometrical interpretation of StimPDC when both the pump and auxiliary lasers are prepared in superpositions of LG modes could be an interesting avenue for future work.

\begin{acknowledgments}
The authors acknowledge financial support from the Brazilian funding agencies Coordena\c{c}\~ao de Aperfei\c{c}oamento de Pessoal de N\'ivel Superior (CAPES), Funda\c{c}\~ao de Amparo \`a Pesquisa do Estado do Rio de Janeiro (FAPERJ), Funda\c{c}\~ao de Amparo \`a Pesquisa do Estado de Alagoas (FAPEAL), and Instituto Nacional de Ci\^encia e Tecnologia de Informa\c{c}\~ao Qu\^antica - Conselho Nacional de Desenvolvimento Cient\'ifico e Tecnol\'ogico (INCT-CNPq).
\end{acknowledgments}
\bibliographystyle{apsrev}
\bibliography{bibliographybib}

\begin{thebibliography}{27}
\expandafter\ifx\csname natexlab\endcsname\relax\def\natexlab#1{#1}\fi
\expandafter\ifx\csname bibnamefont\endcsname\relax
  \def\bibnamefont#1{#1}\fi
\expandafter\ifx\csname bibfnamefont\endcsname\relax
  \def\bibfnamefont#1{#1}\fi
\expandafter\ifx\csname citenamefont\endcsname\relax
  \def\citenamefont#1{#1}\fi
\expandafter\ifx\csname url\endcsname\relax
  \def\url#1{\texttt{#1}}\fi
\expandafter\ifx\csname urlprefix\endcsname\relax\def\urlprefix{URL }\fi
\providecommand{\bibinfo}[2]{#2}
\providecommand{\eprint}[2][]{\url{#2}}

\bibitem[{\citenamefont{Zeilinger}(1999)}]{zeilinger99}
\bibinfo{author}{\bibfnamefont{A.}~\bibnamefont{Zeilinger}},
  \bibinfo{journal}{Rev. Mod. Phys.} \textbf{\bibinfo{volume}{71}},
  \bibinfo{pages}{S288} (\bibinfo{year}{1999}).

\bibitem[{\citenamefont{Walborn et~al.}(2010)\citenamefont{Walborn, Monken,
  P\'{a}dua, and Souto~Ribeiro}}]{walborn10a}
\bibinfo{author}{\bibfnamefont{S.~P.} \bibnamefont{Walborn}},
  \bibinfo{author}{\bibfnamefont{C.~H.} \bibnamefont{Monken}},
  \bibinfo{author}{\bibfnamefont{S.}~\bibnamefont{P\'{a}dua}},
  \bibnamefont{and} \bibinfo{author}{\bibfnamefont{P.~H.}
  \bibnamefont{Souto~Ribeiro}}, \bibinfo{journal}{Phys. Rep.}
  \textbf{\bibinfo{volume}{495}}, \bibinfo{pages}{87} (\bibinfo{year}{2010}).

\bibitem[{\citenamefont{Pan et~al.}(2012)\citenamefont{Pan, Chen, Lu,
  Weinfurter, Zeilinger, and Zukowski}}]{pan12}
\bibinfo{author}{\bibfnamefont{J.-W.} \bibnamefont{Pan}},
  \bibinfo{author}{\bibfnamefont{Z.-B.} \bibnamefont{Chen}},
  \bibinfo{author}{\bibfnamefont{C.-Y.} \bibnamefont{Lu}},
  \bibinfo{author}{\bibfnamefont{H.}~\bibnamefont{Weinfurter}},
  \bibinfo{author}{\bibfnamefont{A.}~\bibnamefont{Zeilinger}},
  \bibnamefont{and} \bibinfo{author}{\bibfnamefont{M.}~\bibnamefont{Zukowski}},
  \bibinfo{journal}{Rev. Mod. Phys.} \textbf{\bibinfo{volume}{84}},
  \bibinfo{pages}{S288} (\bibinfo{year}{2012}).

\bibitem[{\citenamefont{Walther et~al.}(2005)\citenamefont{Walther, Resch,
  Rudolph, Schenck, Weinfurter, Vedral, Aspelmeyer, and Zeilinger}}]{walther05}
\bibinfo{author}{\bibfnamefont{P.}~\bibnamefont{Walther}},
  \bibinfo{author}{\bibfnamefont{K.~J.} \bibnamefont{Resch}},
  \bibinfo{author}{\bibfnamefont{T.}~\bibnamefont{Rudolph}},
  \bibinfo{author}{\bibfnamefont{E.}~\bibnamefont{Schenck}},
  \bibinfo{author}{\bibfnamefont{H.}~\bibnamefont{Weinfurter}},
  \bibinfo{author}{\bibfnamefont{V.}~\bibnamefont{Vedral}},
  \bibinfo{author}{\bibfnamefont{M.}~\bibnamefont{Aspelmeyer}},
  \bibnamefont{and}
  \bibinfo{author}{\bibfnamefont{A.}~\bibnamefont{Zeilinger}},
  \bibinfo{journal}{Nature} \textbf{\bibinfo{volume}{434}},
  \bibinfo{pages}{169} (\bibinfo{year}{2005}).

\bibitem[{\citenamefont{Saleh and Teich}(1991)}]{saleh91}
\bibinfo{author}{\bibfnamefont{B.~E.~A.} \bibnamefont{Saleh}} \bibnamefont{and}
  \bibinfo{author}{\bibfnamefont{M.~C.} \bibnamefont{Teich}},
  \emph{\bibinfo{title}{Fundamentals of Photonics}}
  (\bibinfo{publisher}{Wiley}, \bibinfo{address}{New York},
  \bibinfo{year}{1991}).

\bibitem[{\citenamefont{Zou et~al.}(1991)\citenamefont{Zou, Wang, and
  Mandel}}]{mandel91}
\bibinfo{author}{\bibfnamefont{W.}~\bibnamefont{Zou}},
  \bibinfo{author}{\bibfnamefont{L.}~\bibnamefont{Wang}}, \bibnamefont{and}
  \bibinfo{author}{\bibfnamefont{L.}~\bibnamefont{Mandel}},
  \bibinfo{journal}{Opt. Comm.} \textbf{\bibinfo{volume}{84}},
  \bibinfo{pages}{351} (\bibinfo{year}{1991}).

\bibitem[{\citenamefont{Kwiat et~al.}(1995)\citenamefont{Kwiat, Mattle,
  Weinfurter, Zeilinger, Sergienko, and Shih}}]{kwiat95}
\bibinfo{author}{\bibfnamefont{P.~G.} \bibnamefont{Kwiat}},
  \bibinfo{author}{\bibfnamefont{K.}~\bibnamefont{Mattle}},
  \bibinfo{author}{\bibfnamefont{H.}~\bibnamefont{Weinfurter}},
  \bibinfo{author}{\bibfnamefont{A.}~\bibnamefont{Zeilinger}},
  \bibinfo{author}{\bibfnamefont{A.~V.} \bibnamefont{Sergienko}},
  \bibnamefont{and} \bibinfo{author}{\bibfnamefont{Y.}~\bibnamefont{Shih}},
  \bibinfo{journal}{Phys. Rev. Lett.} \textbf{\bibinfo{volume}{75}},
  \bibinfo{pages}{4337} (\bibinfo{year}{1995}).

\bibitem[{\citenamefont{Kwiat et~al.}(1999)\citenamefont{Kwiat, Waks, White,
  Appelbaum, and Eberhard}}]{kwiat99}
\bibinfo{author}{\bibfnamefont{P.~G.} \bibnamefont{Kwiat}},
  \bibinfo{author}{\bibfnamefont{E.}~\bibnamefont{Waks}},
  \bibinfo{author}{\bibfnamefont{A.~G.} \bibnamefont{White}},
  \bibinfo{author}{\bibfnamefont{I.}~\bibnamefont{Appelbaum}},
  \bibnamefont{and} \bibinfo{author}{\bibfnamefont{P.~H.}
  \bibnamefont{Eberhard}}, \bibinfo{journal}{Physical Review A}
  \textbf{\bibinfo{volume}{60}}, \bibinfo{pages}{R773} (\bibinfo{year}{1999}).

\bibitem[{\citenamefont{Heidmann et~al.}(1987)\citenamefont{Heidmann, Horowicz,
  Reynaud, Giacobino, Fabre, and Camy}}]{heidmann87}
\bibinfo{author}{\bibfnamefont{A.}~\bibnamefont{Heidmann}},
  \bibinfo{author}{\bibfnamefont{R.~J.} \bibnamefont{Horowicz}},
  \bibinfo{author}{\bibfnamefont{S.}~\bibnamefont{Reynaud}},
  \bibinfo{author}{\bibfnamefont{E.}~\bibnamefont{Giacobino}},
  \bibinfo{author}{\bibfnamefont{C.}~\bibnamefont{Fabre}}, \bibnamefont{and}
  \bibinfo{author}{\bibfnamefont{G.}~\bibnamefont{Camy}},
  \bibinfo{journal}{Phys. Rev. Lett.} \textbf{\bibinfo{volume}{59}},
  \bibinfo{pages}{2555} (\bibinfo{year}{1987}).

\bibitem[{\citenamefont{Villar et~al.}(2005)\citenamefont{Villar, Cruz,
  Cassemiro, Martinelli, and Nussenzveig}}]{villar05}
\bibinfo{author}{\bibfnamefont{A.}~\bibnamefont{Villar}},
  \bibinfo{author}{\bibfnamefont{L.}~\bibnamefont{Cruz}},
  \bibinfo{author}{\bibfnamefont{K.}~\bibnamefont{Cassemiro}},
  \bibinfo{author}{\bibfnamefont{M.}~\bibnamefont{Martinelli}},
  \bibnamefont{and}
  \bibinfo{author}{\bibfnamefont{P.}~\bibnamefont{Nussenzveig}},
  \bibinfo{journal}{Physical review letters} \textbf{\bibinfo{volume}{95}},
  \bibinfo{pages}{243603} (\bibinfo{year}{2005}).

\bibitem[{\citenamefont{Barbosa et~al.}(2018)\citenamefont{Barbosa, Coelho,
  Mu{\~n}oz-Mart{\'\i}nez, Ortiz-Guti{\'e}rrez, Villar, Nussenzveig, and
  Martinelli}}]{barbosa18}
\bibinfo{author}{\bibfnamefont{F.}~\bibnamefont{Barbosa}},
  \bibinfo{author}{\bibfnamefont{A.}~\bibnamefont{Coelho}},
  \bibinfo{author}{\bibfnamefont{L.}~\bibnamefont{Mu{\~n}oz-Mart{\'\i}nez}},
  \bibinfo{author}{\bibfnamefont{L.}~\bibnamefont{Ortiz-Guti{\'e}rrez}},
  \bibinfo{author}{\bibfnamefont{A.}~\bibnamefont{Villar}},
  \bibinfo{author}{\bibfnamefont{P.}~\bibnamefont{Nussenzveig}},
  \bibnamefont{and}
  \bibinfo{author}{\bibfnamefont{M.}~\bibnamefont{Martinelli}},
  \bibinfo{journal}{Physical Review Letters} \textbf{\bibinfo{volume}{121}},
  \bibinfo{pages}{073601} (\bibinfo{year}{2018}).

\bibitem[{\citenamefont{Kimble}(2008)}]{kimble08}
\bibinfo{author}{\bibfnamefont{H.~J.} \bibnamefont{Kimble}},
  \bibinfo{journal}{Nature} \textbf{\bibinfo{volume}{453}},
  \bibinfo{pages}{1023} (\bibinfo{year}{2008}).

\bibitem[{\citenamefont{Wang et~al.}(1991)\citenamefont{Wang, Zou, and
  Mandel}}]{wang91}
\bibinfo{author}{\bibfnamefont{L.~J.} \bibnamefont{Wang}},
  \bibinfo{author}{\bibfnamefont{X.~Y.} \bibnamefont{Zou}}, \bibnamefont{and}
  \bibinfo{author}{\bibfnamefont{L.}~\bibnamefont{Mandel}},
  \bibinfo{journal}{J.O.S.A. B} \textbf{\bibinfo{volume}{5}},
  \bibinfo{pages}{978} (\bibinfo{year}{1991}).

\bibitem[{\citenamefont{Souto~Ribeiro et~al.}(1999)\citenamefont{Souto~Ribeiro,
  Padua, and Monken}}]{SoutoRibeiro99}
\bibinfo{author}{\bibfnamefont{P.~H.} \bibnamefont{Souto~Ribeiro}},
  \bibinfo{author}{\bibfnamefont{S.}~\bibnamefont{Padua}}, \bibnamefont{and}
  \bibinfo{author}{\bibfnamefont{C.~H.} \bibnamefont{Monken}},
  \bibinfo{journal}{Phys. Rev. A} \textbf{\bibinfo{volume}{60}},
  \bibinfo{pages}{5074} (\bibinfo{year}{1999}).

\bibitem[{\citenamefont{Parigi et~al.}(2007)\citenamefont{Parigi, Zavatta, Kim,
  and Bellini}}]{parigi07}
\bibinfo{author}{\bibfnamefont{V.}~\bibnamefont{Parigi}},
  \bibinfo{author}{\bibfnamefont{A.}~\bibnamefont{Zavatta}},
  \bibinfo{author}{\bibfnamefont{M.~S.} \bibnamefont{Kim}}, \bibnamefont{and}
  \bibinfo{author}{\bibfnamefont{M.}~\bibnamefont{Bellini}},
  \bibinfo{journal}{Science} \textbf{\bibinfo{volume}{317}},
  \bibinfo{pages}{1890} (\bibinfo{year}{2007}).

\bibitem[{\citenamefont{Caetano et~al.}(2002)\citenamefont{Caetano, Almeida,
  Souto~Ribeiro, Huguenin, Coutinho~dos Santos, and Khoury}}]{Caetano02}
\bibinfo{author}{\bibfnamefont{D.~P.} \bibnamefont{Caetano}},
  \bibinfo{author}{\bibfnamefont{M.~P.} \bibnamefont{Almeida}},
  \bibinfo{author}{\bibfnamefont{P.~H.} \bibnamefont{Souto~Ribeiro}},
  \bibinfo{author}{\bibfnamefont{J.~A.~O.} \bibnamefont{Huguenin}},
  \bibinfo{author}{\bibfnamefont{B.}~\bibnamefont{Coutinho~dos Santos}},
  \bibnamefont{and} \bibinfo{author}{\bibfnamefont{A.~Z.}
  \bibnamefont{Khoury}}, \bibinfo{journal}{Phys. Rev. A}
  \textbf{\bibinfo{volume}{66}}, \bibinfo{pages}{041801(R)}
  (\bibinfo{year}{2002}).

\bibitem[{\citenamefont{Caetano et~al.}(2001)\citenamefont{Caetano, Almeida,
  Souto~Ribeiro, Huguenin, Coutinho~dos Santos, and Khoury}}]{caetano01}
\bibinfo{author}{\bibfnamefont{D.~P.} \bibnamefont{Caetano}},
  \bibinfo{author}{\bibfnamefont{M.~P.} \bibnamefont{Almeida}},
  \bibinfo{author}{\bibfnamefont{P.~H.} \bibnamefont{Souto~Ribeiro}},
  \bibinfo{author}{\bibfnamefont{J.~A.~O.} \bibnamefont{Huguenin}},
  \bibinfo{author}{\bibfnamefont{B.}~\bibnamefont{Coutinho~dos Santos}},
  \bibnamefont{and} \bibinfo{author}{\bibfnamefont{A.~Z.}
  \bibnamefont{Khoury}}, \bibinfo{journal}{Phys. Rev. Lett.}
  \textbf{\bibinfo{volume}{87}}, \bibinfo{pages}{133602}
  (\bibinfo{year}{2001}).

\bibitem[{\citenamefont{Padget and Courtial}(1999)}]{padgett99}
\bibinfo{author}{\bibfnamefont{M.~J.} \bibnamefont{Padget}} \bibnamefont{and}
  \bibinfo{author}{\bibfnamefont{J.}~\bibnamefont{Courtial}},
  \bibinfo{journal}{Opt. Lett.} \textbf{\bibinfo{volume}{24}},
  \bibinfo{pages}{430} (\bibinfo{year}{1999}).

\bibitem[{\citenamefont{Coutinho~dos Santos
  et~al.}(2007)\citenamefont{Coutinho~dos Santos, Souza, Dechoum, and
  Khoury}}]{santos07}
\bibinfo{author}{\bibfnamefont{B.}~\bibnamefont{Coutinho~dos Santos}},
  \bibinfo{author}{\bibfnamefont{C.~E.~R.} \bibnamefont{Souza}},
  \bibinfo{author}{\bibfnamefont{K.}~\bibnamefont{Dechoum}}, \bibnamefont{and}
  \bibinfo{author}{\bibfnamefont{A.~Z.} \bibnamefont{Khoury}},
  \bibinfo{journal}{Phys. Rev. A} \textbf{\bibinfo{volume}{76}},
  \bibinfo{pages}{053821} (\bibinfo{year}{2007}).

\bibitem[{\citenamefont{Rodrigues et~al.}(2018)\citenamefont{Rodrigues,
  Gonzales, Pinheiro~da Silva, Huguenin, Martinelli, Medeiros~de Ara\'ujo,
  Souza, and Khoury}}]{rodrigues2018}
\bibinfo{author}{\bibfnamefont{R.~B.} \bibnamefont{Rodrigues}},
  \bibinfo{author}{\bibfnamefont{J.}~\bibnamefont{Gonzales}},
  \bibinfo{author}{\bibfnamefont{B.}~\bibnamefont{Pinheiro~da Silva}},
  \bibinfo{author}{\bibfnamefont{J.~A.~O.} \bibnamefont{Huguenin}},
  \bibinfo{author}{\bibfnamefont{M.}~\bibnamefont{Martinelli}},
  \bibinfo{author}{\bibfnamefont{R.}~\bibnamefont{Medeiros~de Ara\'ujo}},
  \bibinfo{author}{\bibfnamefont{C.~E.~R.} \bibnamefont{Souza}},
  \bibnamefont{and} \bibinfo{author}{\bibfnamefont{A.~Z.}
  \bibnamefont{Khoury}}, \bibinfo{journal}{Optics Letters}
  \textbf{\bibinfo{volume}{43}}, \bibinfo{pages}{2486} (\bibinfo{year}{2018}).

\bibitem[{\citenamefont{Vaity et~al.}(2013)\citenamefont{Vaity, Banerji, and
  Singh}}]{vaity2013}
\bibinfo{author}{\bibfnamefont{P.}~\bibnamefont{Vaity}},
  \bibinfo{author}{\bibfnamefont{J.}~\bibnamefont{Banerji}}, \bibnamefont{and}
  \bibinfo{author}{\bibfnamefont{R.~P.} \bibnamefont{Singh}},
  \bibinfo{journal}{Physics letters A} \textbf{\bibinfo{volume}{377}},
  \bibinfo{pages}{1154} (\bibinfo{year}{2013}).

\bibitem[{\citenamefont{Mair et~al.}(2001)\citenamefont{Mair, Vaziri, Weihs,
  and Zeilinger}}]{Mair2001}
\bibinfo{author}{\bibfnamefont{A.}~\bibnamefont{Mair}},
  \bibinfo{author}{\bibfnamefont{A.}~\bibnamefont{Vaziri}},
  \bibinfo{author}{\bibfnamefont{G.}~\bibnamefont{Weihs}}, \bibnamefont{and}
  \bibinfo{author}{\bibfnamefont{A.}~\bibnamefont{Zeilinger}},
  \bibinfo{journal}{Nature} \textbf{\bibinfo{volume}{412}},
  \bibinfo{pages}{313} (\bibinfo{year}{2001}).

\bibitem[{\citenamefont{Walborn et~al.}(2004)\citenamefont{Walborn,
  de~Oliveira, Thebaldi, and Monken}}]{Walborn04a}
\bibinfo{author}{\bibfnamefont{S.~P.} \bibnamefont{Walborn}},
  \bibinfo{author}{\bibfnamefont{A.~N.} \bibnamefont{de~Oliveira}},
  \bibinfo{author}{\bibfnamefont{R.~S.} \bibnamefont{Thebaldi}},
  \bibnamefont{and} \bibinfo{author}{\bibfnamefont{C.~H.}
  \bibnamefont{Monken}}, \bibinfo{journal}{Phys. Rev. A}
  \textbf{\bibinfo{volume}{69}}, \bibinfo{pages}{023811}
  (\bibinfo{year}{2004}).

\bibitem[{\citenamefont{Walborn et~al.}(2005)\citenamefont{Walborn, P{\'a}dua,
  and Monken}}]{Walborn2005}
\bibinfo{author}{\bibfnamefont{S.}~\bibnamefont{Walborn}},
  \bibinfo{author}{\bibfnamefont{S.}~\bibnamefont{P{\'a}dua}},
  \bibnamefont{and} \bibinfo{author}{\bibfnamefont{C.}~\bibnamefont{Monken}},
  \bibinfo{journal}{Physical Review A} \textbf{\bibinfo{volume}{71}},
  \bibinfo{pages}{053812} (\bibinfo{year}{2005}).

\bibitem[{\citenamefont{Straupe et~al.}(2011)\citenamefont{Straupe, Ivanov,
  Kalinkin, Bobrov, and Kulik}}]{Straupe2011}
\bibinfo{author}{\bibfnamefont{S.}~\bibnamefont{Straupe}},
  \bibinfo{author}{\bibfnamefont{D.}~\bibnamefont{Ivanov}},
  \bibinfo{author}{\bibfnamefont{A.}~\bibnamefont{Kalinkin}},
  \bibinfo{author}{\bibfnamefont{I.}~\bibnamefont{Bobrov}}, \bibnamefont{and}
  \bibinfo{author}{\bibfnamefont{S.}~\bibnamefont{Kulik}},
  \bibinfo{journal}{Physical Review A} \textbf{\bibinfo{volume}{83}},
  \bibinfo{pages}{060302} (\bibinfo{year}{2011}).

\bibitem[{\citenamefont{Arruda et~al.}(2018)\citenamefont{Arruda, Soares,
  Walborn, Tasca, Kanaan, Medeiros~de Ara\'ujo, and
  Souto~Ribeiro}}]{Arruda2018}
\bibinfo{author}{\bibfnamefont{M.~F.~Z.} \bibnamefont{Arruda}},
  \bibinfo{author}{\bibfnamefont{W.~C.} \bibnamefont{Soares}},
  \bibinfo{author}{\bibfnamefont{S.~P.} \bibnamefont{Walborn}},
  \bibinfo{author}{\bibfnamefont{D.~S.} \bibnamefont{Tasca}},
  \bibinfo{author}{\bibfnamefont{A.}~\bibnamefont{Kanaan}},
  \bibinfo{author}{\bibfnamefont{R.}~\bibnamefont{Medeiros~de Ara\'ujo}},
  \bibnamefont{and} \bibinfo{author}{\bibfnamefont{P.~H.}
  \bibnamefont{Souto~Ribeiro}}, \bibinfo{journal}{Phys. Rev. A}
  \textbf{\bibinfo{volume}{98}}, \bibinfo{pages}{023850}
  (\bibinfo{year}{2018}).

\bibitem[{\citenamefont{Silva and Khoury}(2018)}]{Silva2018}
\bibinfo{author}{\bibfnamefont{M.~P.} \bibnamefont{Silva}} \bibnamefont{and}
  \bibinfo{author}{\bibfnamefont{A.~Z.} \bibnamefont{Khoury}},
  \bibinfo{journal}{to be submitted}  (\bibinfo{year}{2018}).

\end{thebibliography}

\end{document}